\documentclass{article}

\usepackage{arxiv}

\usepackage[round]{natbib} 
\usepackage[utf8]{inputenc} 
\usepackage[T1]{fontenc}    
\usepackage{hyperref}       
\usepackage{url}            
\usepackage{booktabs}       
\usepackage{amsfonts}       
\usepackage{nicefrac}       
\usepackage{microtype}      
\usepackage{graphicx}
\usepackage{amsmath}
\usepackage{array}
\newcolumntype{L}[1]{>{\raggedright\arraybackslash}p{#1}}
\usepackage{tabularx}
\usepackage{multirow}  
\usepackage{xcolor}
\usepackage{float}
\usepackage{placeins}  
\usepackage{cleveref}
\usepackage[normalem]{ulem}
\useunder{\uline}{\ul}{}
\usepackage{colortbl}
\usepackage{subcaption}
\usepackage{enumitem}

\title{Expanding External Access to Frontier AI Models for Dangerous Capability Evaluations}

\author{
 Jacob Charnock \\
  ERA Cambridge\\
  Cambridge, UK \\
  \texttt{jakecharnock25@gmail.com} \\
   \And
 Alejandro Tlaie \\
  Pour Demain\\
  Brussels, Belgium \\
  \phantom{\texttt{jakecharnock25@gmail.com}} \\
  \And
 Kyle O'Brien \\
  ERA Cambridge\\
  Cambridge, UK \\
  \phantom{\texttt{jakecharnock25@gmail.com}} \\
  \AND
 Stephen Casper \\
  MIT CSAIL\\
  Cambridge, MA, USA \\
  \And
 Aidan Homewood \\
  GovAI\\
  London, UK \\
}

\begin{document}
\maketitle
\begin{abstract}
Frontier AI companies increasingly rely on external evaluations to assess risks from dangerous capabilities before deployment. However, external evaluators often receive limited model access, limited information, and little time, which can reduce evaluation rigour and confidence. The EU General-Purpose AI Code of Practice calls for ``appropriate'' access, but does not specify what this means in practice. Furthermore, there is no common framework for describing different types and levels of evaluator access. To address this gap, we propose a taxonomy of access methods for dangerous capability evaluations. We disentangle three aspects of access: model access, model information, and evaluation timeframe. For each aspect, we review benefits and risks, including how expanding access can reduce false negatives and improve stakeholder trust, but can also increase security and capacity challenges. We argue that these limitations can likely be mitigated through technical means and safeguards used in other industries. Based on the taxonomy, we propose three descriptive access levels: AL1 (black-box model access and minimal information), AL2 (grey-box model access and substantial information), and AL3 (white-box model access and comprehensive information), to support clearer communication between evaluators, frontier AI companies, and policymakers. We believe these levels correspond to the different standards for appropriate access defined in the EU Code of Practice, though these standards may change over time.
\end{abstract}

\newpage

\section*{Executive Summary}

This paper provides a taxonomy for describing external access for model evaluations, and offers a preliminary set of Access Levels corresponding to the different standards for appropriate access in the EU Code of Practice (CoP). 

\subsection*{The current state of external evaluator access (\Cref{sec:introduction})}
Frontier AI companies increasingly rely on external evaluation results to assess whether their models are safe for deployment. However, external evaluators are typically given limited model access, technical information, and time to complete their evaluations. The CoP outlines high-level expectations for ``appropriate'' access for third-party model evaluations; however, it does not clarify what this means in practice. 

\subsection*{The relationship between access, rigour, and security (\Cref{sec:relationship})}
Expanding external evaluator access has two key benefits. First, expanding access can make external evaluations more rigorous, helping to reduce false negatives (i.e. cases where dangerous capabilities exist but remain undetected) as well as false positives (i.e. cases where evaluators overestimate dangerous capabilities). Second, expanding external access can increase stakeholder trust (e.g. amongst governments, regulators, and customers) because evaluators can assess systems with greater independence, and in more breadth and depth. Despite this, expanding evaluator access, absent appropriate safeguards, can present challenges. Whilst unlikely, third parties with enough access could leak proprietary information. Furthermore, providing additional external access could strain internal capacity at frontier companies. Nevertheless, we find these risks can likely be mitigated through technical means (e.g. privacy-preserving access), and mitigations used in other industries (e.g. segregating duties, appointing a liaison). 

\subsection*{A taxonomy of structured access for dangerous capability evaluations (\Cref{sec:taxonomy})}
Our taxonomy disentangles three distinct aspects of access: 
\\
\\
\textbf{Model access} can range from black-box querying to full white-box access including activations, gradients, logits, and custom fine-tuning. Deeper model access enables more accurate capability assessments by enabling different evaluation techniques (e.g. using log-probabilities to reduce variance). However, more model access could increase opportunities for model tampering, reverse engineering, and model weight theft unless appropriate safeguards are in place. 
\\
\\
\textbf{Model information} can include information about training data, deployment configurations, internal evaluation results, and safety mitigation details. Greater information disclosure helps evaluators contextualise their results, identify blind spots, probe potential vulnerabilities, and more accurately verify company safety claims. Despite this, more information disclosure risks exposing proprietary design and training choices that competitors could steal or adversaries could exploit, unless strong information handling protocols are in place. 
\\
\\
\textbf{Evaluation time frames} can vary significantly for external reviewers, ranging from no access before release, less than a week pre-release, to assessments exceeding 20 business days. Longer evaluation time frames enable more thorough evaluations, as evaluators can design bespoke testing methods, re-run, adapt and debug evaluations. However, longer time frames impose greater operational costs on developers (e.g. allocating internal employees to oversee longer external evaluations) and could interfere with model release schedules. 

\subsection*{Access Levels (\Cref{sec:accesslevels})}
Based on the taxonomy described in (\Cref{sec:taxonomy}), we outline three descriptive access levels. We relied on existing literature, expert discussions, and definitions from the CoP to categorise the access methods from our taxonomy. We believe these access levels correspond to each definition of ``appropriate'' access under the EU GPAI Code of Practice: AL1 (i.e. black-box model access and minimal information), AL2 (i.e. grey-box model access and substantial information), and AL3 (i.e. white-box model access and comprehensive information).

\begin{figure}[H]
\centering
\begin{tabular}{p{14cm}}
\toprule
\textbf{AL1: Black-box model access and minimal information (``best practice'')} \\[2mm]
This level enables basic external assurance and vulnerability detection. It might be suitable when a developer releases a minor update to an existing model, or when a new model poses low systemic risk. \\
\midrule
\textbf{AL2: Grey-box model access and substantial information (``the state of the art'')} \\[2mm]
This level enables more targeted evaluations to improve capability elicitation and vulnerability identification. It might be suitable for model releases with significant capability jumps. \\
\midrule
\textbf{AL3: White-box model access and comprehensive information (``more innovative'')} \\[2mm]
This level likely enables the most accurate risk assessments. It might be suitable for models with step-changes in capabilities, when model deployment decisions depend on evaluation results, or if privacy-preserving tools can support white-box access methods. \\
\bottomrule
\end{tabular}
\label{fig:accesslevels}
\end{figure}

\clearpage

\section{Introduction} \label{sec:introduction}
Frontier AI companies increasingly use results from external model evaluations to demonstrate to stakeholders and governments that their models are sufficiently safe for deployment \citep{METR2025, FrontierModelForum2025}. External evaluations of the dangerous capabilities of frontier AI models have become part of risk assessments before deployment \citep{GPT-5system,claude4_5system,Googlemodelcard}. Results from these evaluations can provide governments and the public with more confidence in the safety of frontier AI models, allowing them to make more informed decisions \citep{Bommasani2023, Staufer2025}. Similarly, evaluation results can help companies improve their safeguards \citep{WorkwithCAISI,antrhopicsafeguards}, and help governments identify and anticipate emerging risks \citep{AISIo1}. External evaluations are included in the Frontier AI Safety Commitments \citep{DSIT2024}, and the EU General-Purpose AI Code of Practice (“CoP”) \citep{EuropeanCommission2025}.

However, external evaluators often receive limited access to the models they are evaluating, which reduces evaluation rigour and confidence \cite{Casper2024, Staufer2025}. External evaluators are often restricted to black-box access, meaning they can query the model and examine outputs (sometimes alongside chains-of-thought), but cannot examine any model internals (e.g. activations, gradients, weights). Furthermore, external evaluators often receive limited technical information, which means they often struggle to fully assess reported properties \citep{Casper2024} and provide stakeholders with high confidence evaluation results \citep{OpenMined2023}. Moreover, despite an improvement in external access for recent evaluations involving chain-of-thought (CoT) access and longer evaluation time frames \citep{METR2025GPT-5, GPT-5system}, external evaluators continue to receive inconsistent model access, limited (often unverifiable) information about models, and days rather than weeks to evaluate models.\footnote{For example, Apollo and UK AISI received less than a week to evaluate Claude Sonnet 4.5 \citep{claude4_5system}.} Some researchers have proposed “structured access” arrangements (i.e. controlled model access through interfaces that enable expanded access while protecting against misuse) to facilitate better external access for evaluations \citep{Shevlane2022,Bucknall2023,Casper2024,Bucknall2025}. Privacy-preserving access tools, such as secure enclaves \citep{Trask2024} and zero-knowledge proofs \citep{Waiwitlikhit2024, South2024}, could facilitate deeper model access without exposing sensitive information \citep{Tlaie2025}. Structured access approaches could also facilitate secure access to model information, such as training datasets \citep{Waiwitlikhit2024}. However, before being deployed reliably at scale, these tools must likely become more efficient on large-scale computations \citep{Gamiz2025} and less operationally burdensome.\footnote{For example, secure enclaves still require all parties to be online simultaneously \citep{Trask2024}.}

Regulatory and governance frameworks covering model access for external evaluations have recently emerged. The CoP outlines requirements for Signatories on providing external evaluators with “appropriate” access to models, information about these models, and time to complete their evaluations \citep{EuropeanCommission2025}. According to the CoP, what is “appropriate” may either be “best practice”, “the state of the art”, or “other more innovative processes”. Another framework is the AEF-1 standard, which outlines “minimum operating conditions” for external evaluations, including a requirement for “sufficient technical access to assess the specific system characteristics being evaluated” \citep{Stosz2025}. While these documents indicate access methods that might be sufficient, they do not describe concretely when each method is necessary, nor do they outline the full range of possible access methods and their respective benefits and risks. 

While prior work outlines some of the ways external evaluator access could be improved \citep{Casper2024, Bucknall2025, Stosz2025}, this paper provides a taxonomy of different forms of external access for dangerous capability evaluations. Our taxonomy aims to help companies, evaluators, and policymakers communicate more clearly about what forms of access are given or expected across key dimensions and improve the quality of external model evaluations. First, we consider how expanding external access could improve AI evaluations (\Cref{sec:relationship}). Second, we outline our taxonomy for expanding external access – disentangling model access, model information, and evaluation time frame (\Cref{sec:taxonomy}). Third, we propose three access levels based on our taxonomy: AL1 (black-box model access and minimal information), AL2 (grey-box model access and substantial information), and AL3 (white-box access and comprehensive information) (\Cref{sec:accesslevels}). We believe these levels correspond to the different standards for “appropriate” access defined in the CoP as (1) the current “best practice”, (2) the current “state of the art”, and (3) “more innovative” access methods. 

\section{The Relationship Between Access, Rigour, and Security} \label{sec:relationship}

This section outlines how expanding external access can enable more rigorous and meaningful evaluations of frontier models. We outline the main benefits and challenges associated with improving external access for dangerous capability evaluations, and explore some potential mitigations. 

\subsection{Benefits of Expanding External Access} \label{sec:benefits}

There are two key benefits to frontier AI companies when they provide external evaluators with more access. 

\textbf{More rigorous evaluations.} Expanding external access could significantly enhance evaluation quality and reliability. Providing external evaluators with model information such as internal evaluation results, methodologies, and safeguard details, could help expose internal blind spots \citep{Longpre2025, 4opus} and poorly calibrated thresholds in dangerous capability assessments. This is because external reviewers also bring diverse expertise and methods that can enhance evaluation coverage and accuracy \citep{Longpre2025}. Providing deeper model access could reveal latent vulnerabilities and dangerous capabilities \citep{Casper2024, TF2025}. Access to the internal model state (e.g. activations) enables external evaluators to red-team the system more effectively, achieving more robust evaluation results than black-box queries alone \citep{Che2025}. Expanding external access could therefore help reduce false negatives (i.e. cases where dangerous capabilities exist but remain undetected). This is important because if developers better understand model capabilities, they can make better deployment decisions \citep[see][]{AnthropicAL3}. Expanded access including more time and resources could also reduce false positives (i.e. cases where evaluators overestimate a model’s dangerous capabilities) as evaluators can re-run more tests and have more context to run better evaluations, ensuring that model risks are not overstated.

\textbf{Greater stakeholder confidence.} Expanding external access can enhance the verifiability of evaluation outcomes and increase external trust (e.g. from governments, regulators, and the public) in company safety claims \citep{Casper2024}. For example, if third parties had access to sampling controls, and the evaluation methods used by frontier AI companies (e.g. whether evaluations were conducted on fine-tuned variants), they could more accurately interpret \citep{Bucknall2023, Biderman2024} and validate reported results \citep{Casper2024}. Additionally, stable external access to models (i.e. sustained access to the same model despite being superceded by new updates) is important for the independent replication of evaluation runs \citep{Reuel2025} and helps ensure impartiality as external reviewers do not have a clear conflict of interest \citep{TF2025, Longpre2025}, and developers cannot selectively disclose or cherry-pick high-performance runs \citep{Roque2024, Singh2025}. Access to more resources and model artifacts also allows evaluators to assess systems in greater breadth and depth, enabling better-supported external evaluation results that can strengthen stakeholder trust. 

\subsection{Challenges of Expanding External Access} \label{sec:challenges}

There are two key challenges frontier AI companies face when providing external evaluators with additional access.

\textbf{Security risks.} Depending upon how access is provided, expanding external access to frontier models can increase the risks of IP and security leaks for a frontier AI company. For example, providing evaluators with the ability to fine-tune models and access safety classifiers would allow them to obtain security-critical information about harmful capabilities, bypass strategies, and classifier vulnerabilities that could aid adversaries if leaked, be deliberately misused by evaluators, or cause reputational damage if concerning model behaviours became public. With grey-box access, evaluators may also be able to reveal sensitive IP such as architectural details (e.g. hidden dimensions) – for example, by analysing log-probabilities returned by the model to reverse-engineer internal details like the model's size and architecture \citep{Carlini2024}. In extreme cases evaluators could also use their additional access to misuse models. For instance, if a malicious evaluator wanted help conducting a cyberattack, they could use a helpful-only version of the model to assist them. Beyond model access, there are security risks when third parties gain access to high-level information about security architecture \citep{Benaroch2021}. Moreover, evaluators that conduct reviews for multiple AI companies could inadvertently transfer tacit knowledge about systems to other frontier AI companies raising potential IP and security concerns \citep{Kang2021}. 

However, frontier AI companies could mitigate security risks by using technical, legal, and physical safeguards \citep{Casper2024}. One technical safeguard for access to models is using structured access methods. These methods can allow evaluators to analyse systems using some grey and white-box tools through an application programming interface (API), without providing evaluators direct access to model parameters \citep{Casper2024, Bucknall2023}. Technical safeguards for access to model information include secure document management and transfer tools (e.g. encrypted file transfers or secure collaboration platforms) to safely transfer sensitive materials \citep{Jagadeeswari2023}, and audit trails to record who connected to servers and what they did (e.g. files accessed, operations performed) \citep{Duncan2016, ISO27002}. Legal safeguards include non-disclosure agreements, engagement letters, and established rules of engagement between the evaluator and frontier AI company \citep{NIST2008}. In other industries, companies can protect IP and prevent security critical information leaks by agreeing to access controls. Examples of access controls include providing auditors with temporary accounts that automatically disable, least privilege access so evaluators are limited to the information necessary for their role, separation of duties, and prohibiting wireless network connections on non-company devices \citep{NIST2020}. If physical access is needed, security mechanisms include on-site document review rooms, escort protocols, personnel screening and event reports \citep{Lightman2022, ISO27002, ISO27001}, which are routinely used in other industries \citep{FinancialConductAuthority2024, OfficeoftheComptrolleroftheCurrency2018, MedicinesandHealthcareproductsRegulatoryAgency2025, USFoodDrugAdministration2024}. 

\textbf{Capacity constraints.} Expanding external access for evaluations may constrain internal capacity within a frontier AI company. As noted above, frontier AI companies may need to use technical, legal, and physical safeguards to prevent security risks associated with expanding access. These safeguards will require staff time and experience to implement. However, frontier AI companies may not have personnel with experience implementing these safeguards, and those with experience may not have the time to do so.

There are at least two approaches to mitigating internal capacity constraints. First, frontier AI companies and evaluators can follow emerging standards to have greater clarity on what protocols should be followed for the evaluation, and spend less time negotiating terms \citep[see][]{Stosz2025}. Second, companies could draw upon practices used in third-party assessments in other industries, such as appointing a dedicated liaison to manage communication and coordinate the evaluation \citep[see][]{Ismael2018}. 

The two challenges above can likely be mitigated through techniques used in other industries. Since companies often voluntarily undertake external assessments that are not required by law \citep{Trustcentre,Trustportal}, the benefits from expanding access for external evaluations might therefore outweigh the IP, security, and internal capacity challenges they create.

\section{A Taxonomy of Structured Access for Dangerous Capability Evaluations} \label{sec:taxonomy}

In this section, we disentangle three key aspects of expanding external access for dangerous capability evaluations: Model access (Section 3.1); Model information (Section 3.2); and Evaluation time frame (Section 3.3). These aspects align with the requirement in the CoP for Signatories to provide evaluators with adequate resources \citep{EuropeanCommission2025}. To determine the access methods for each aspect of the taxonomy, we first examined current company practices by surveying publicly available evaluation reports by frontier AI companies and evaluators \citep{GPT-5system,claude4_5system, METR2025GPT-5}, information about previously provided access methods that are available to the public \citep{OPENPERM}, and existing standards \citep{Stosz2025} and recommendations for evaluator access \citep{Bucknall2025}. We then examined more innovative practices by surveying literature on open-weight model evaluations \citep{Che2025,Schwinn2024,Sadasivan2024} and other reviews of the evaluation field \citep{TF2025,Casper2024}. To categorise methods within each aspect, we relied on existing literature \citep{Casper2024} and expert discussions. For each aspect, we outline plausible access options and discuss their main benefits and limitations. We also consider interdependencies across these dimensions (Section 3.4). 

\subsection{Model Access} \label{sec:modelaccess}

External evaluators could be given various types of model access to evaluate a model for dangerous capabilities. We classify these types of access into three levels of model access: Black-box access (i.e. access to query the system and observe outputs); grey-box access (i.e. limited access to model internals such as input embeddings, hidden neural activations, or log-probabilities); and white-box access (i.e. full access to the system including access to activations, gradients, and the ability to fine-tune the model) \citep{Casper2024}. \Cref{tab:modelaccess} sets out these levels of model access.

\begin{table}[htbp]
\small
\centering
\caption{Overview of different depths of model access.}
\vspace{3mm}
\label{tab:modelaccess}
\renewcommand{\arraystretch}{1.2}
\begin{tabular}{@{}p{3.5cm}p{10.5cm}@{}}
\toprule
\textbf{Depth of model access} & \textbf{Model access methods} \\
\midrule
No external access before deployment & 
\vspace{-2mm}\begin{itemize}[noitemsep,topsep=0pt,leftmargin=*]
\item None
\end{itemize}\vspace{-2mm} \\
\midrule
Black-box access & 
\vspace{-2mm}\begin{itemize}[noitemsep,topsep=0pt,leftmargin=*]
\item Black-box access pre-release (i.e.\ ability to query the model and examine model outputs).
\item Model checkpoints (i.e.\ sample model versions from various points during training)
\item Helpful-only model (i.e.\ sample model versions without harmlessness training)
\item Raw CoT (i.e.\ view model reasoning traces)
\item Enable/disable input prompt safety classifiers (i.e.\ ability to disable input/output classifiers when sampling from models)
\end{itemize}\vspace{-2mm} \\
\midrule
Grey-box access & 
\vspace{-2mm}\begin{itemize}[noitemsep,topsep=0pt,leftmargin=*]
\item Unfiltered fine-tuning (i.e.\ ability to fine-tune the model on custom data)
\item Change sampling algorithm (i.e.\ ability to select from different algorithms and adjust parameters that determine how tokens are selected)
\item Log-probabilities (i.e.\ view the log probabilities of output tokens)
\item Logit access (i.e.\ view the unnormalised scores the model assigns to each possible next token)
\item Input embeddings, inner neuron activations (i.e.\ ability to observe how inputs are encoded and transformed as they pass through model layers)
\item Access to input and output classifier responses (i.e.\ view the responses of safety filters to prompts before they reach the model, and outputs before they reach the user)
\end{itemize}\vspace{-2mm} \\
\midrule
White-box access & 
\vspace{-2mm}\begin{itemize}[noitemsep,topsep=0pt,leftmargin=*]
\item Full access to activations, gradients, and logits (i.e.\ ability to view and change the model's internal states, training signals, and pre-decoded outputs)
\item Fine-tune with custom loss function (i.e.\ ability to specify a loss function that defines how model parameters are updated)
\item Full access to input/output classifiers (i.e.\ ability to directly inspect and test safety classifiers separately from the model).
\end{itemize}\vspace{-2mm} \\
\bottomrule
\end{tabular}
\end{table}

\textbf{No external access before deployment.} This means that evaluators cannot query or interact with the model pre-release, and are limited to testing the commercially available model rather than through a special API. The main benefit of providing no external access pre-release is that it minimises security risks and capacity constraints. However, evaluators cannot reliably evaluate models because deployment level guardrails can obscure true capability ceilings and rate-limits may prevent statistically significant evaluation results. More importantly, evaluators cannot detect dangerous capabilities prior to release, meaning harms could accrue due to companies making worse deployment decisions.

\textbf{Black-box access.} This level allows evaluators to query the model and examine model outputs (sometimes alongside chains-of-thought), but not examine any model internals (e.g. weights, activations, gradients). The main benefit of this level is that it enables evaluators to identify dangerous capabilities, jailbreaks, or unsafe propensities before deployment whilst maintaining strong security guarantees. Evaluators can also gain a better understanding of potential upper bounds on dangerous capabilities and estimate safety filter protections (e.g. by querying helpful-only models without refusals, and disabling classifiers). Chain-of-thought (CoT) access can help distinguish refusals and apparent underperformance from true capability ceilings and detect dangerous reasoning patterns \citep{METR2025GPT-5}. However, black-box access limits the ability of evaluators to obtain high-confidence results because sampling alone cannot efficiently cover the vast input space, meaning results are sensitive to which elicitation techniques the evaluator uses. Moreover, black-box methods have proved unreliable for detecting jailbreaks and backdoors \citep{Casper2024}. 

\textbf{Grey-box access.} This level gives third parties limited access to classifier responses, internal model states (including logits, log-probabilities, and inner neuron activations) and the ability to fine-tune on custom data. This level of access can improve the accuracy of evaluation results. Access to logits and log-probabilities can allow for more accurate estimates of capabilities and propensities by enabling evaluators to directly estimate performance distributions, rather than relying on black-box sampling \citep{Bucknall2025,Miller2024}. Fine-tuning on custom data helps elicit harmful latent behaviours, giving a more reliable upper-bound on capability estimates \citep{Casper2024,Qi2023}.\footnote{Fine-tuning access is especially valuable for evaluating models that downstream users can modify (e.g. open-weight models \citep{OBrien2025, Wallace2025}), or for models that users can fine-tune via an API (e.g. GPT-4o fine-tuning on Azure OpenAI Service).} Grey-box access also allows evaluators to use more red-teaming algorithms (e.g. latent-space attacks) which can more accurately detect latent behaviours and vulnerabilities missed by black-box sampling \citep{Che2025,Schwinn2024}. However, this level of access brings more significant security risks as evaluators may be able to reveal sensitive IP such as architectural details (e.g. hidden dimensions) by analysing log-probabilities returned to reverse-engineer internal details like the model's size and architecture \citep{Carlini2024}. Moreover, details from fine-tuning (e.g. custom datasets and bypass strategies) or stress-testing classifiers (e.g. vulnerabilities discovered) could aid misuse if improperly stored or leaked. 

\textbf{White-box access.} This level gives evaluators full access to logits, activations, gradients, and classifiers, alongside the ability to fine-tune with a custom loss function. This level enables the most accurate evaluations as white-box methods substantially expand the option space for evaluators \cite{Casper2024}. For instance, white-box model attacks (e.g. latent-space interventions, model-tampering attacks) can help uncover dangerous capabilities or misaligned behaviours entirely missed under input-output probing \citep{Che2025,Hofsttter2025}. Additionally, evaluators can inspect internal mechanisms (e.g. activations, attention heads, residual stream) via gradient-based probes or interpretability tools to reveal internal representations (e.g. “lying,” “honesty,”) \citep{Zou2025}, and latent capabilities independently of deployment \citep{TF2025}. Fine-tuning with custom loss functions further supports evaluators in testing for dangerous latent capabilities or hidden behaviours \citep{Casper2024, Bucknall2023}. Classifier access also allows evaluators to craft more targeted tests that systematically probe safeguard weaknesses \citep{WorkwithCAISI,antrhopicsafeguards}. However, this white-box access carries the most significant security risks because providing evaluators with access to activations and gradients increases their ability to reverse engineer the model, potentially enabling them to recover internal parameters \citep{Milli2018}, embedding layers \citep{Carlini2024} or even weights \citep{Zanella-Bguelin2021}. In practice, however, the risks of providing white-box or very light grey-box access can be reduced through APIs, where model parameters do not leave the developers' servers, or on-site evaluations \citep{Casper2024}.

Overall, deeper model access enables higher-confidence evaluations, but increases the risk of IP and security related information leaks \citep{Casper2024,TF2025,Bommasani2023,Nevo2024}. To accommodate deeper levels of access, frontier AI companies and evaluators should implement technical, legal, and physical safeguards.

\subsection{Model Information} \label{sec:modelinfo}

Evaluators could be given various types of non-public model information to support their dangerous capability evaluations. We classify these types of information into four main groups: Training data, deployment configurations, internal evaluation processes and results, and safety mitigations. For each type of information disclosure we outline minimal, substantial and comprehensive levels of disclosure. \Cref{tab:modelinfo} sets out these different levels of model information disclosure.

\begin{table}[htbp]
\small
\centering
\caption{Minimal, substantial, and comprehensive levels of disclosure for each of the four types of information.}
\vspace{3mm}
\label{tab:modelinfo}
\renewcommand{\arraystretch}{1.2}
\begin{tabular}{@{}L{3cm}p{2.6cm}p{8.4cm}@{}}
\toprule
\textbf{Type of information} & \textbf{Level of disclosure} & \textbf{Information provided} \\
\midrule
\textbf{Training data}
& Minimal & 
\vspace{-2mm}\begin{itemize}[noitemsep,topsep=0pt,leftmargin=*]
\item Excluded topics \& websites/sources
\item Included dual-use topics
\item Model specification
\item Domains covered in fine-tuning and safety training
\end{itemize}\vspace{-2mm}\\
& Substantial & 
\vspace{-2mm}\begin{itemize}[noitemsep,topsep=0pt,leftmargin=*]
\item Proportions of dataset that contain dual-use information
\item Estimates of dual-use data that made it past filters
\item Prevalence of dual-use topics by language and modality
\end{itemize}\vspace{-2mm}\\
& Comprehensive & 
\vspace{-2mm}\begin{itemize}[noitemsep,topsep=0pt,leftmargin=*]
\item Quantitative dataset composition (i.e.\ proportions in tracked categories, language coverage percentages, real/synthetic split)
\item Data curation pipeline (i.e.\ human data contractors, inclusion/exclusion criteria, filtering, sampling/deduplication criteria)
\item Synthetic data pipeline (i.e.\ any generation methods used, safety measures taken to protect generation, methods for validating safety of data curation)
\item Controlled querying of datasets
\end{itemize}\vspace{-2mm}\\
\midrule
\textbf{Deployment configurations}
& Minimal & 
\vspace{-2mm}\begin{itemize}[noitemsep,topsep=0pt,leftmargin=*]
\item System prompt
\item Model affordances
\end{itemize}\vspace{-2mm}\\
& Substantial & 
\vspace{-2mm}\begin{itemize}[noitemsep,topsep=0pt,leftmargin=*]
\item Model context protocols (including non-public configurations for tools, data access, and applications)
\item Inference-time settings (e.g.\ sampling parameters)
\end{itemize}\vspace{-2mm}\\
& Comprehensive & 
\vspace{-2mm}\begin{itemize}[noitemsep,topsep=0pt,leftmargin=*]
\item Architecture of agent scaffolding, including safety steps towards filtering external knowledge sources
\end{itemize}\vspace{-2mm}\\
\midrule
\textbf{Internal evaluation processes and results}
& Minimal & 
\vspace{-2mm}\begin{itemize}[noitemsep,topsep=0pt,leftmargin=*]
\item Assertions about internal evaluation runs including conflicting results, dangerous propensities, elicitation techniques
\end{itemize}\vspace{-2mm}\\
& Substantial & 
\vspace{-2mm}\begin{itemize}[noitemsep,topsep=0pt,leftmargin=*]
\item Description of evaluation methodology, frequency, confidence intervals to accompany reported results
\item Evaluation sample size, and percentage of distribution covered by evaluation set
\end{itemize}\vspace{-2mm}\\
& Comprehensive & 
\vspace{-2mm}\begin{itemize}[noitemsep,topsep=0pt,leftmargin=*]
\item Full internal evaluation reports (methods, thresholds, and results)
\item Red-teaming transcripts
\item Safeguard efficacy analysis results
\item Acceptable use policies
\end{itemize}\vspace{-2mm}\\
\midrule
\textbf{Safety mitigations}
& Minimal & 
\vspace{-2mm}\begin{itemize}[noitemsep,topsep=0pt,leftmargin=*]
\item List of implemented safety mitigations (e.g.\ monitoring/filtering inputs and outputs, fine-tuning for refusals, anti-tampering safeguards, staged access strategies (i.e.\ for roll out and roll back))
\item List of types of incidents tracked and known vulnerabilities
\end{itemize}\vspace{-2mm}\\
& Substantial & 
\vspace{-2mm}\begin{itemize}[noitemsep,topsep=0pt,leftmargin=*]
\item Classification metrics and filter thresholds for safeguards (CoT Monitors, input/output Classifiers)
\item Techniques for achieving CoT faithfulness
\item Architecture of filters, monitors, and refusal logic
\item Quantitative performance metrics of safety filters
\end{itemize}\vspace{-2mm}\\
& Comprehensive & 
\vspace{-2mm}\begin{itemize}[noitemsep,topsep=0pt,leftmargin=*]
\item Direct access to inference-time auxiliary classifiers/safety scaffolding (e.g. input/output classifiers, chain-of-thought filtering, content moderation models)
\item Safety data sets for refusals (incl. adversarial prompt sets).
\item Internal safety mitigations (e.g.\ employee access provisions)
\end{itemize}\vspace{-2mm}\\
\bottomrule
\end{tabular}
\end{table}

\textbf{Training data.} Disclosing training data helps evaluators infer what dangerous capabilities models may have learned, and assess data-related safety interventions, but increases intellectual property and security risks. If the company prioritises minimising security risks over the accuracy of evaluations, they may provide high-level information about training data via the model specification (a document outlining the model’s intended behaviour and design choices \citep{OpenAI2025}), alongside details of excluded topics, websites, included dual-use topics, and fine-tuning domains to provide a rough indication of a model’s capabilities \citep{Udandarao2024, Longpre2024} (minimal). If the company weighs the accuracy of evaluations and protecting security similarly, they may disclose more detailed proportions of datasets (e.g. proportions of dual-use data) to help evaluators anticipate learned capabilities and design targeted vulnerability tests \citep{Longpre2024} (substantial). If the company prioritises the accuracy of evaluations over security risks, or companies can mitigate these risks, companies may disclose a detailed breakdown of the contents of their datasets, examples from fine-tuning datasets, details about data curation practices, and safety precautions taken during training (comprehensive). However, comprehensive disclosure risks exposing proprietary dataset curation methods and safety training that competitors could replicate \citep{Bucknall2025, Casper2023} or adversaries could exploit.

\textbf{Deployment configurations.} Disclosing deployment configurations can help evaluators understand the risks a model may pose by clarifying the deployed system’s affordances, but can increase security and IP risks if sensitive configurations are leaked. If companies prioritise minimising information security risks over enabling more accurate evaluations, they may provide the model system prompt, and model affordances (i.e. permissions, tools, and scaffolds) to help evaluators contextualise and improve the trustworthiness of assessments \citep{Stosz2025} (minimal).\footnote{For example, evaluators may identify if a failure to answer dangerous questions is due to a lack of capability or a safety constraint in the system prompt, and more accurately evaluate whether models adhere to stated behavioural guidelines \citep{Ahmed2025}.} If companies weigh more accurate evaluations and information security risks similarly, they may reveal non-public details about model context protocols (e.g. configurations for tools, data access, and applications) \citep{MCP} and inference-time settings (e.g. sampling parameters) so evaluators better understand system affordances and can more accurately assess risks the deployed model might pose (substantial). If companies prioritise more accurate evaluations over information exposure to evaluators, or can mitigate information sharing risks, they may share the architecture of agent scaffolding (e.g. prompt construction, and interaction loop management) as well as safety steps for filtering external knowledge sources (comprehensive). This could help evaluators assess system capabilities, since scaffolding can significantly affect performance even when the underlying model is unchanged \citep{SWE}. They may also test whether retrieval mechanisms could be exploited. However, comprehensive disclosure risks exposing sensitive IP related to performance and details that attackers could exploit if leaked. Companies could mitigate this trade-off by providing access using servers that automatically disable access after pre-agreed periods.

\textbf{Internal evaluation processes and results.} Disclosing internal model evaluation processes and results helps external evaluators contextualise their own findings and more accurately interpret and verify company safety claims, but increases information security risks and operational burden. If companies prioritise reducing information security risk over providing confirming external evaluation results, they may provide high-level information about internal evaluation runs (e.g. through yes/no attestations) to help contextualise and improve evaluator judgements (minimal).\footnote{METR \citep{METR2025GPT-5} received an attestation from OpenAI that their findings did not contradict OpenAI’s internal results.} If they weigh confirming external evaluation results and information security risk similarly, they could also provide details about evaluation methodology to help evaluators interpret results, identify limitations \citep{McCaslin2025} and increase stakeholder trust in safety claims \citep{Staufer2025,Bommasani2024} (substantial). If companies value confirming external evaluation results over minimising information security risk, or they can mitigate the risks of sharing evaluation results effectively, they might also share full transcripts of internal evaluations and safeguard efficacy analysis results to help evaluators verify reported results, and identify incorrect methods or misleading results \citep{McCaslin2025, Bowen2025}.\footnote{This includes cases where developers only report results with safeguards applied, making systems appear less dangerous than they might be if jailbreaks are discovered in deployment \citep{Bowen2025}. Full evaluation reports and transcripts could reveal whether poor performance reflects genuine capability limits or inadequate testing \citep{McCaslin2025,StevenAdler2025,Bowen2025} and help distinguish genuine alignment behaviours from memorisation or evaluation awareness – a concern highlighted by Apollo over Claude Sonnet 4.5’s low deception rates \citep{claude4_5system}.} Despite this, more detailed evaluation transcripts might reveal hazardous information (e.g. information about specific pathogens, scaffolding methods, or safeguard vulnerabilities) that could cause harm if misused by evaluators, or leaked \citep{McCaslin2025}. Frontier AI companies could alleviate this trade-off by redacting sensitive information that is not core to evaluation results \citep{Review}, restricting high-risk reports (e.g. biological content policies) to external evaluators with strong security measures, or using non-disclosure agreements (NDAs) to prevent information sharing. 

\textbf{Safety mitigations.} Disclosing safety mitigations helps evaluators assess the robustness of deployment safeguards and identify vulnerabilities, but increases information security risks that might aid adversarial exploitation. If companies prioritise minimising information exposure over assurances about their safety mitigations, they may provide high-level lists of implemented safety mechanisms, and categories of tracked incidents (minimal). This provides evaluators with a general understanding of safety approaches and tracked areas, while limiting information that could facilitate bypass attempts. If companies weigh assurances about their safety mitigations and information exposure risks similarly, they may also disclose classification metrics and filters for safeguards (e.g. precision/recall for CoT monitors and input/output classifiers), thresholds for interventions, architectural details of filters and refusal logic, and quantitative performance metrics. These artefacts should help evaluators identify gaps in safety coverage and verify that monitoring systems are appropriately robust (substantial) \citep[see][]{WorkwithCAISI, antrhopicsafeguards}. If companies prioritise providing assurances about their safety mitigations over minimising information exposure, they may also disclose filter thresholds, classifier weights, safety training datasets and internal safety mitigations \citep[e.g. employee access provisions, see][]{METR2025Transparency} to enable more targeted testing of whether safety mitigations and organisational safety processes are adequate for mitigating systemic risks (comprehensive). However, comprehensive disclosure increases the risk of leaking precise implementation details (e.g. adversarial prompt sets) that could help adversaries develop more effective attacks that undermine safety protections. Companies could mitigate this trade-off through controlled access environments where evaluators test safeguards and view filters or thresholds under monitored conditions, including time-limited access, least-privileged access, and audit trails. 

\subsection{Evaluation Time Frame} \label{sec:timeframe}
Another key aspect of external evaluations is the time frame that evaluators receive to conduct their evaluations. The CoP states that at least 20 business days is appropriate for most model evaluations \citep{EuropeanCommission2025}. In practice, however, evaluation time frames are often much shorter than this. For example, external evaluators for Claude Sonnet 4 and Claude Sonnet 4.5 received only 1 week of access (see \citealt{4opus} and \citealt{claude4_5system}). Below we consider the benefits and challenges of shorter and longer evaluation time frames. Given that the CoP suggests that external evaluations require adequate staffing, we assume that third-party staffing is constant across time frames, so that longer access corresponds to more expert hours. 

\textbf{Shorter evaluation time frames.} When external evaluators receive shorter pre-release access to a model (e.g. one to three weeks) they can often only identify obvious vulnerabilities. With short evaluation time frames, evaluators can use basic prompting and red-teaming methods to uncover obvious vulnerabilities or elicit some capabilities before deployment. These approaches can surface potential vulnerabilities and rough capability thresholds relatively quickly by relying on existing tools such as shared prompt engineering resources, and public benchmarks. However, under these shorter time frames (e.g. ~ three weeks or less), evaluators usually lack the time to design new evaluation methods, adapt existing ones, build bespoke testing scaffolds, or re-run failed tests \citep[see][]{METRo3}. This means capability elicitation can be shallow and implies that clearer conclusions could be reached with longer evaluation time frames \citep{USAISIUKAISI2024}. Key benefits of short evaluation time frames include that they are convenient for developers to meet model release deadlines, do not require dedicated internal capacity for long periods, and can provide some confidence to stakeholders that model capability thresholds are within acceptable bounds pre-release. Key challenges include that capability evaluations lack depth and evaluator confidence in results is low because there is insufficient time to design robust evaluations or properly elicit and scaffold model behaviour.

\textbf{Longer evaluation time frames.} By contrast, providing longer evaluation time frames enables evaluators to properly design, adapt, debug, and execute their evaluation methods. To this end, the CoP requires Signatories provide evaluators with at least 20 business days for most model evaluations \citep{EuropeanCommission2025}. Extended access enables more thorough capability elicitation, including by creating bespoke evaluations, and gaining the opportunity to resample evaluations for statistical robustness. This means longer evaluation time frames increase the likelihood of uncovering additional vulnerabilities or hidden capabilities. The main benefit of longer evaluation time frames is that evaluators can develop and run more comprehensive evaluations which provides more accurate and higher confidence results. In turn, this strengthens stakeholder and public trust, because safety claims can be validated by independent third parties that received adequate time and resources to conduct thorough evaluations. The main challenges of providing longer evaluation time frames are that they require greater oversight from frontier AI companies and may delay model releases. Models may also undergo significant changes during the 20 business day period before release, meaning that the model version which evaluators receive may be too different from the model that is eventually deployed. 

Overall, the time frame of an evaluation is a key component for improving the comprehensiveness and accuracy of evaluations. There is a trade-off then between the operational costs of providing access for a longer period of time and the level of assurance that evaluators can offer regarding a model’s safety. 

\subsection{Interdependencies} \label{sec:interdependencies}
These different aspects of external evaluations can depend upon each other in important ways. While the preferences of companies might determine the level of access they provide for each aspect, choices for one aspect can inform or constrain what is feasible for another. For example: 

\begin{itemize}
    \item If a company imposes short evaluation periods, the value of providing substantial or extensive model information diminishes because evaluators are unlikely to be able to review and make use of additional information.
    \item If a company is concerned about the IP risks of disclosing substantial model information, they could provide deeper model access to external evaluators to support better results, without significant information disclosure. 
    \item If a company cannot afford to give evaluators longer evaluation periods due to internal capacity constraints and deadlines, this could be partially offset by providing either richer model information which can enable faster contextualisation of results, or deeper model access (e.g. observing logits or log-probabilities) which allows evaluators to achieve more statistically significant results across fewer samples due to a reduction in the variance of evaluation results \citep{Bucknall2025}.
    \item If a company offers deep access in one dimension (e.g. grey-box access), it may be sensible elsewhere to provide complementary forms of access that would improve evaluation results, whilst adding little additional security or IP risk (e.g. substantive model information). 
\end{itemize}

Importantly, different aspects of access do not appear to be substitutable. For instance, some vulnerabilities (e.g. backdoors, hidden representation biases) will be easier to detect with white-box access but very hard to detect with black box access, even if evaluators have access to the full dataset and model specification. Similarly, evaluators typically need more time to make use of additional forms of model access. While evaluation aspects are interdependent, stakeholders should use our taxonomy as a menu to choose the different levels of access to provide that suit their use case.

\section{Access Levels} \label{sec:accesslevels}

Below we propose three access levels. Access Level 1 (AL1) refers to black-box model access and minimal information. Access Level 2 (AL2) refers to grey-box model access and substantial information. Access Level 3 (AL3) refers to white-box access and comprehensive information. We relied on existing literature \citep{Casper2024}, expert discussions, and definitions from the CoP to categorise the access methods outlined in our taxonomy into three access levels. The CoP describes three standards that could be used to meet “appropriate”: “best practice” (i.e. accepted practices among model providers as techniques that best assess and mitigate systemic risk); “the state of the art” (i.e. the forefront of relevant research, governance, and technology that goes beyond best practice); and “more innovative” techniques (i.e. techniques more advanced than the state of the art) \citep{EuropeanCommission2025}. Importantly, using “best practice” or “state of the art” techniques is not necessarily sufficient to comply with the CoP. We believe our access levels correspond to the different standards for appropriate access defined in the CoP. These access levels will require corrections and adjustments over time because companies and external evaluators will change their practices.

\subsection{Access Level 1} \label{sec:al1}

Access Level 1 (AL1) enables basic external assurance and vulnerability detection, and is the minimum access level appropriate for frontier AI models under the CoP. This might be suitable for external evaluations of dangerous capabilities when developers release minor updates to an existing model, or when a new model poses low systemic risk. AL1 allows external evaluators to provide no more than basic assurance because evaluators only receive black-box access with limited information disclosures and minimal time. This means that evaluators will likely only identify obvious vulnerabilities as they are restricted to basic sampling techniques, limited information about the model, and insufficient time to re-run or design bespoke evaluations.

\begin{table}[htbp]
\small
\centering
\caption{Overview of Access Level 1}
\vspace{3mm}
\label{tab:al1}
\renewcommand{\arraystretch}{1.2}
\begin{tabular}{@{}L{3.2cm}L{2.8cm}L{7.8cm}@{}}
\toprule
\textbf{Access Dimension} & \textbf{Subcategory} & \textbf{Access provided} \\
\midrule
Model access & -- & 
\vspace{-2mm}\begin{itemize}[noitemsep,topsep=0pt,leftmargin=*]
\item Black-box API
\item Model checkpoints
\item Helpful-only variants
\item CoT
\item Ability to enable/disable input output classifiers
\end{itemize}\vspace{-2mm} \\
\midrule
Model information & Training data & 
\vspace{-2mm}\begin{itemize}[noitemsep,topsep=0pt,leftmargin=*]
\item Excluded topics/sources
\item Included dual-use topics
\item Model specification
\item Domains covered in fine-tuning and safety training
\end{itemize}\vspace{-2mm} \\
\\
& Deployment configurations & 
\vspace{-2mm}\begin{itemize}[noitemsep,topsep=0pt,leftmargin=*]
\item System prompt
\item Model affordances
\end{itemize}\vspace{-2mm} \\
\\
& Internal evaluation processes and results & 
\vspace{-2mm}\begin{itemize}[noitemsep,topsep=0pt,leftmargin=*]
\item Assertions about internal evaluation runs including conflicting results, dangerous propensities, elicitation techniques
\end{itemize}\vspace{-2mm} \\
\\
& Safety mitigations & 
\vspace{-2mm}\begin{itemize}[noitemsep,topsep=0pt,leftmargin=*]
\item List of implemented safety mitigations
\item Classification metrics for safeguards
\item Filter/monitor architecture and refusal logic
\item Performance metrics of safety filters
\end{itemize}\vspace{-2mm} \\
\midrule
Evaluation time frame & -- & 
\vspace{-2mm}\begin{itemize}[noitemsep,topsep=0pt,leftmargin=*]
\item At least 20 business days
\end{itemize}\vspace{-2mm} \\
\bottomrule
\end{tabular}
\end{table}

AL1 corresponds to the CoP's “best practice” definition of appropriate access, and establishes the minimum access Signatories need to provide to external evaluators under the CoP \citep{EuropeanCommission2025}. Signatories likely need to provide evaluators with black-box access to the model before release, because this appears to be best practice \citep{GPT-5system,claude4_5system, Amazon2025}. Pursuant to Measure 3.5 of the CoP, Signatories need to provide external evaluators access to model versions with the fewest safety mitigations, alongside the chains-of-thought, for external post-market monitoring. This suggests Signatories likely need to provide model checkpoints, variants, and the chain-of-thought during external evaluations for full risk assessments. This is proportional because the security risks and operational costs of providing access to model versions without safety measures and their CoTs are likely similar for external post-market monitoring and full risk assessment. Pursuant to Appendix 3.3 of the CoP, external evaluators need to assess the effectiveness of mitigations, which suggests evaluators need substantial information about safety mitigations, alongside which dual-use topics were included in the training data, and domains covered by safety training. Pursuant to Appendix 3.4 of the CoP, if appropriate for the systemic risk and model evaluation method, Signatories must provide evaluation teams with the model spec, system prompt, training data, test sets, and past model evaluation results. Evaluators may also need this information to elicit the model, pursuant to Appendix 3.2 of the CoP. The AEF-1 standard recommends external evaluators gain access to system prompts, information about the training process and data, pre-existing internal evaluation results, and knowledge of system vulnerabilities as relevant for trustworthy evaluations, among other forms of information \citep{Stosz2025}. Pursuant to Appendix 3.4 of the CoP, providing evaluators at least 20 business days to conduct evaluations is appropriate for most systemic risks (presumably CBRN, loss of control, cyber offense, and harmful manipulation) and evaluation methods (presumably red teaming, capability elicitation, scaffolded evaluations, and bespoke task suites).

\subsection{Access Level 2} \label{sec:al2}

Access Level 2 (AL2) allows more targeted evaluations to improve capability elicitation and vulnerability identification. This might be suitable for model releases with significant capability jumps. AL2 supports evaluators in providing more targeted and reliable model evaluations through some detailed information about systems, access to sampling controls and fine-tuning, and longer evaluation windows. These affordances mean that evaluators can design and rerun bespoke evaluations, resample from the model for statistical robustness, and use fine tuning and safety classifier information to more accurately stress-test safeguards.

\begin{table}[htbp]
\small
\centering
\caption{Overview of Access Level 2}
\vspace{3mm}
\label{tab:al2}
\renewcommand{\arraystretch}{1.2}
\begin{tabular}{@{}L{3.2cm}L{2.8cm}L{7.8cm}@{}}
\toprule
\textbf{Access Dimension} & \textbf{Subcategory} & \textbf{Access provided} \\
\midrule
Implementation of previous Access Levels & -- & 
\vspace{-2mm}\begin{itemize}[noitemsep,topsep=0pt,leftmargin=*]
\item Developer has implemented all access from AL1
\end{itemize}\vspace{-2mm} \\
\midrule
Model access & -- & 
\vspace{-2mm}\begin{itemize}[noitemsep,topsep=0pt,leftmargin=*]
\item Unfiltered fine-tuning
\item Change sampling algorithm
\item Log-probabilities
\item Classifier responses
\end{itemize}\vspace{-2mm} \\
\midrule
Model information & Training data & 
\vspace{-2mm}\begin{itemize}[noitemsep,topsep=0pt,leftmargin=*]
\item The same as for AL1
\end{itemize}\vspace{-2mm} \\
\\
& Deployment configurations & 
\vspace{-2mm}\begin{itemize}[noitemsep,topsep=0pt,leftmargin=*]
\item Model context protocols
\item Inference-time settings (e.g.\ sampling parameters)
\end{itemize}\vspace{-2mm} \\
\\
& Internal evaluation processes and results & 
\vspace{-2mm}\begin{itemize}[noitemsep,topsep=0pt,leftmargin=*]
\item Detailed description of evaluation methodologies to accompany reported results
\end{itemize}\vspace{-2mm} \\
\\
& Safety mitigations & 
\vspace{-2mm}\begin{itemize}[noitemsep,topsep=0pt,leftmargin=*]
\item Same as AL1
\end{itemize}\vspace{-2mm} \\
\midrule
Evaluation time frame & -- & 
\vspace{-2mm}\begin{itemize}[noitemsep,topsep=0pt,leftmargin=*]
\item $\geq$20 business days
\end{itemize}\vspace{-2mm} \\
\bottomrule
\end{tabular}
\end{table}

AL2 corresponds to the current state of the art for external evaluator access under the CoP. For instance, frontier AI companies have previously provided heightened model access in the form of supervised fine-tuning on custom datasets \citep{OPENPERM}, the ability to specify or modify some sampling parameters \citep{Saunders2022}, and access to view log-probabilities of output tokens \citep{OpenAILogprobs2023}.\footnote{OpenAI, Anthropic, and Google have each previously provided a selection of various sampling algorithms and control of relevant parameters \citep[see][Appendix A]{Bucknall2023}} Whilst substantial training data disclosure could likely be mitigated effectively, and the CoP suggests that relevant training data information for systemic risks and model evaluation methods should be provided, no frontier AI company appears to have disclosed training data beyond that of AL1. Pursuant to Appendix 3.4 of the CoP, past model evaluation results must be provided as appropriate for each systemic risk and model evaluation method. Recent work has provided a template for frontier model providers that allows them to more thoroughly describe their evaluation processes and results without publishing information posing security risks \citep{McCaslin2025, Reed2025}, suggesting that providing evaluators with this level of detail about internal evaluations may be state-of-the-art. Pursuant to Appendix 3.3 of the CoP, external evaluators must assess the effectiveness of safety mitigations at an appropriate breadth and depth, including under adversarial pressure. This suggests that evaluators may need heightened model access (e.g. to fine-tune the model, probe safeguards, or generate jailbreaks) as well as detailed information about safety mitigation and deployment configuration details. OpenAI and Anthropic have previously provided UK AISI and US CAISI access to several versions of constitutional classifiers and additional safety mitigation details (e.g. safeguard architecture details, documented vulnerabilities, content policy information, classifier scores) to more accurately identify safeguard vulnerabilities \citep{WorkwithCAISI, antrhopicsafeguards}. No frontier AI company appears to have provided external evaluators with more time to conduct their evaluations than the minimal requirement of 20 business days pursuant to Appendix 3.4. 

\subsection{Access Level 3} \label{sec:al3}
Access Level 3 (AL3) represents a more experimental access arrangement which likely allows for the most accurate external evaluation results. These affordances mean that evaluators can more rigorously test capability ceilings, identify vulnerabilities, and use non-public information to validate reported safety results. This might be suitable for models with significant jumps in capabilities when model deployment decisions depend on evaluation results, or when privacy-preserving tools can reliably support white-box access. 

\begin{table}[htbp]
\small
\centering
\caption{Overview of Access Level 3.}
\vspace{3mm}
\label{tab:al3}
\renewcommand{\arraystretch}{1.2}
\begin{tabular}{@{}L{3.2cm}L{2.8cm}L{7.8cm}@{}}
\toprule
\textbf{Access Dimension} & \textbf{Subcategory} & \textbf{Access provided} \\
\midrule
Implementation of previous Access Levels & -- & 
\vspace{-2mm}\begin{itemize}[noitemsep,topsep=0pt,leftmargin=*]
\item Developer has implemented all access from AL1 and AL2
\end{itemize}\vspace{-2mm} \\
\midrule
Model access & -- & 
\vspace{-2mm}\begin{itemize}[noitemsep,topsep=0pt,leftmargin=*]
\item Full access to activations, gradients, and logits
\item Ability to fine-tune with custom loss function
\item Full access to input/output classifiers
\end{itemize}\vspace{-2mm} \\
\midrule
Model information & Training data & 
\vspace{-2mm}\begin{itemize}[noitemsep,topsep=0pt,leftmargin=*]
\item Detailed breakdown of the contents of datasets
\item Details on data curation pipeline and safety protections
\end{itemize}\vspace{-2mm} \\
\\
& Deployment configurations & 
\vspace{-2mm}\begin{itemize}[noitemsep,topsep=0pt,leftmargin=*]
\item Architecture of agent scaffolding (incl.\ safety steps towards filtering external knowledge sources)
\end{itemize}\vspace{-2mm} \\
\\
& Internal evaluation processes and results & 
\vspace{-2mm}\begin{itemize}[noitemsep,topsep=0pt,leftmargin=*]
\item Full internal reports and transcripts
\item Safeguard efficacy analysis results
\end{itemize}\vspace{-2mm} \\
\\
& Safety mitigations & 
\vspace{-2mm}\begin{itemize}[noitemsep,topsep=0pt,leftmargin=*]
\item Filter thresholds
\item Safety data sets (incl.\ adversarial prompt datasets)
\item Internal access rules for employees (i.e. for preventing unauthorized system access)
\end{itemize}\vspace{-2mm} \\
\midrule
Evaluation time frame & -- & 
\vspace{-2mm}\begin{itemize}[noitemsep,topsep=0pt,leftmargin=*]
\item Greater than 20 business days
\end{itemize}\vspace{-2mm} \\
\bottomrule
\end{tabular}
\end{table}

AL3 corresponds to access methods that would go beyond the current state of the art for external access under the CoP. Signatories could provide white-box access to models. Since white-box access has proved most useful for uncovering harmful behaviours, vulnerabilities, and potential sandbagging on open-source models \citep{Che2025, Casper2024, Taylor2025}, but has not been piloted by external evaluators on closed source models, this likely surpasses the current state of the art. Pursuant to Appendix 3.2 of the CoP, Signatories may provide additional affordances (e.g. fine-tuning with a custom loss function, or activations) to minimise the risk of under elicitation and undetected sandbagging during model evaluations.\footnote{The AI Evaluator Forum recommends that fine-tuning should be provided in cases where users can finetune the model \citep[see][]{Stosz2025}. Access to activations can be useful for running probes which may help identify model sandbagging behaviours \citep{Taylor2025}.} Pursuant to Appendix 3.4 of the CoP, Signatories must provide evaluation teams with training data, test sets, and past model evaluation results as appropriate for each systemic risk and evaluation method. A more experimental access arrangement might therefore involve providing substantial information, including about data composition and post training details to better understand how training data can affect a model’s behaviour \citep{Soldaini2024}, data curation policies to identify unexamined biases that may cause downstream model behaviours \citep{Dodge2021}, and full internal evaluation reports to validate reported results. Pursuant to Appendix 3.3 of the CoP, Signatories need to assess the extent to which mitigations work as planned and can be circumvented or subverted. In a more experimental setup, frontier AI companies could provide filter thresholds and outline internal safety mitigations (e.g. internal employee access provisions) to help evaluators assess safety mitigations. Finally, a more experimental access arrangement would likely require companies provide external evaluators with more than 20 business days to accommodate more experimental white-box evaluation methods (e.g. latent space attacks) and more information artifacts to study.

In selecting an access level for external evaluations, companies will need to consider whether the risks posed by their model warrant minimal, state of the art, or more experimental access methods. Importantly, companies do not need to provide the same level of access across all dimensions or for different model releases. Moreover, they can adopt access methods that align with their specific priorities and risk profiles. Nevertheless, we think that AL2 is clearly within reach for frontier AI companies today. This is because AL2 is unlikely to pose significant security risks, and any remaining risks can likely be mitigated through existing technical and legal safeguards (e.g. API access, secure enclaves, NDAs), as well as measures commonly used for third-party assessments in other industries (e.g. segregating duties).

\section{Conclusion} \label{sec:conclusion}

In this paper we have outlined a taxonomy of access to promote clearer standards and communication for external dangerous capability evaluations. We have attempted to address the problem of inconsistent reporting by providing a framework for companies and evaluators to communicate about what forms of access might be required across key dimensions (model access, model information, and time), and by proposing “access levels” that correspond to the external access requirements in the EU GPAI Code of Practice. We reviewed the benefits and risks of expanding access, and argued that key challenges can likely be mitigated through technical means and safeguards used in other industries.
There are two main limitations to our work. First, we determined access levels based upon the public information that is available about the current access provided for external evaluations. Therefore, our access levels may not correspond to the access that companies currently provide to external evaluators. This means our judgement of minimal, state of the art, and more innovative access levels may not be the most accurate approximations. Second, the CoP is open to interpretation, meaning that there remains scope for disagreement over which levels of access are appropriate in different situations. 

The paper has also left several questions unanswered that warrant further research. Future work could examine how to use our taxonomy and access levels to make access requirements more concrete in practice. Future work could also clarify the conditions (e.g. the level of risk, or required evaluation method) under which external evaluators need deeper access to a model. Finally, future work could develop tools for structured access, and use these tools in future evaluations, so that external evaluator access can become less costly and more secure.

We view our taxonomy and access levels as a starting point, and expect that they will require updates as risks emerge, evaluation methods change, and tools for structured access improve. We therefore invite researchers, frontier AI companies, evaluators, and regulators to use and iterate on our taxonomy and access levels.

\section*{Acknowledgments}
This research was supported by the ERA Fellowship. The authors would like to thank ERA for their financial and intellectual support. We are grateful for valuable comments and feedback from Alan Chan, Ben Bucknall, Dave Banerjee, Edward Kembery, Harrison Gietz, Jonas Freund, Lisa Soder, Lucas Sato, May Dixit, Nikola Jurkovic, Patricia Paskov, Prakriti Bandhan, Tom Reed, Zaheed Kara, and all the participants of the ERA 2025 Summer Fellowship. Any remaining errors are our own.

\bibliographystyle{plainnat}
\bibliography{refs}

\end{document}